\documentclass[print]{aa} 

\usepackage{graphicx}
\usepackage{txfonts}
\usepackage{longtable}
\usepackage{natbib}
\bibpunct{(}{)}{;}{a}{}{,}

\begin{document}

   \title{Catalogue of ISO\thanks{Based on observations with ISO, an ESA project
          with instruments funded by ESA Member States (especially
          the PI countries: France, Germany, the Netherlands
          and the United Kingdom) and with the participation
          of ISAS and NASA.} LWS observations of asteroids}

   \author{F. Hormuth
          \inst{1,2}
          \and
          T. G. M\"uller\inst{3}}

   \offprints{F.\ Hormuth}

   \institute{Centro Astron\'omico Hispano Alem\'an,
		C/ Jes\'us Durb\'an Rem\'on 2-2$^\circ$, 04004 Almer\'ia, Spain
      \and 
   Max-Planck-Institut f\"ur Astronomie (MPIA),
              K\"onigstuhl 17, 69117 Heidelberg, Germany\\
              \email{hormuth@mpia.de}
       \and
             Max-Planck-Institut f\"ur extraterrestrische Physik (MPE),
             Giessenbachstrasse, 85748 Garching, Germany\\
             \email{tmueller@mpe.mpg.de}
             }

   \date{Received; accepted}


  \abstract
   {The Long Wavelength Spectrometer (LWS) onboard the Infrared Space 
   Observatory (ISO) observed the four large main-belt asteroids \object{(1)~Ceres}, 
   \object{(2)~Pallas}, \object{(4)~Vesta}, and \object{(10)~Hygiea} multiple times. 
   The photometric and spectroscopic data cover the wavelength range between 
   43 and 197\,$\mu$m, and are a unique dataset for future investigations and 
   detailed characterisations of these bodies.}
   {The standard ISO archive products, produced through the last post-mission
   LWS pipeline, were still affected by instrument artefacts. 
   Our goal was to provide the best possible data products to exploit the full 
   scientific potential of these observations.}
   {For all asteroid observations we analysed in detail the dark
   current, the calibration reference flashes, the space environment
   effects (glitches), memory effects, tracking influences,
   and various other sources of uncertainty. 
   We performed a refined reduction of all measurements, corrected for the various 
   effects, and re-calibrated the data. 
   We outline the data reduction process and  give an overview of the available data 
   and the quality of the observations. 
   We apply a thermophysical model to the flux measurements to derive far-IR 
   based diameter and albedo values of the asteroids.
   The measured thermal rotational lightcurve of (4)~Vesta is compared to model predictions.  }
   {The catalogue of LWS (Long Wavelength Spectrometer) observations
   of asteroids contains 57 manually reduced datasets, including
   seven non-standard observations, which as such did not have final
   pipeline products available before. 
   In total, the archive now contains 11 spectral scans and 46 fixed grating 
   measurements with a simultaneous observation at 10 key wavelengths 
   distributed over the full LWS range. 
   The new data products are now accessible via the ISO data archive as highly 
   processed data products (HPDP)\footnote{\texttt{http://iso.esac.esa.int/ida/hpdp.php}}.
   }
   {The quality of the data products was checked against state-of-the-art
   thermophysical model predictions and an excellent agreement was found.
   The absolute photometric accuracy is better than 10\,\%. The calibrated
   spectra will serve as source for future mineralogical studies of dwarf
   planets and dwarf planet candidates. }


   \maketitle
%

\section{Introduction}

The Infrared Space Observatory \citep[ISO; ][]{Kessler:1996}
observed between 1995 and 1998 more than 40 asteroids in great detail,
including some complete spectra from 2 to 200\,$\mu$m and large samples
of photometric measurements \citep{Muller:2003,Dotto:2002}. 
The main goals of the about 100\,hours of asteroid observing time were
the identification of surface minerals, composition, connection to meteorites
and comets, surface alteration processes, and the interpretation of taxonomic
classes through the identification of mid-infrared features of well-known
minerals and meteorites. 
\citet{Muller:2005} summarised the main ISO results from the various solar 
system programmes and from all four ISO instruments.

The ISO Long Wavelength Spectrometer \citep[LWS;][]{Clegg:1996} 
performed photometric and spectroscopic measurements in the wavelength
range 43 to 197\,$\mu$m and opened, together with ISOPHOT, a new window 
for far-infrared astronomy.
The details on the instrument and the data processing can be found
in the ISO Handbook, Volume III \citep{Gry:2003}.
In total, the ISO Data Archive (IDA\footnote{\tt http://www.iso.esac.esa.int/ida/index.html})
contains 51 successful LWS observations of four different asteroids
covering almost 18\,hours of satellite time. 
However, the scientific return of these programmes was so far very little. 
Most of the scientific questions could not be answered mainly due to the 
instrumental artefacts  which remained in the archived data products. 

The LWS data set of asteroid observations covers a wavelength range which
is not accessible from the ground. 
In the near future the Herschel mission will be capable of obtaining asteroid spectra
between 57 and 210\,$\mu$m with PACS (Photodetector Array Camera \& Spectrometer)
and between 200 and 670\,$\mu$m with SPIRE (Spectral and Photometric Imaging Receiver),
but asteroids are currently not part of the accepted key programmes.
Although the ISO target list is very limited, it includes objects of high
scientific interest: 
(1)~Ceres and (4)~Vesta are the targets of the Dawn\footnote{\tt http://dawn.jpl.nasa.gov/} 
spacecraft mission, (1)~Ceres is meanwhile considered as a dwarf-planet, and (2)~Pallas, 
(4)~Vesta, and (10)~Hygiea are dwarf-planet candidates\footnote{Dissertatio Cvm Nvncio 
Sidereo III, Nr. 3, August 16, 2006, {\tt http://astro.cas.cz/nuncius/}}. 
In general, it is believed that these largest bodies in the main-belt are 
protoplanets remaining intact since their formation.
 
The LWS observations comprise spectral energy distributions over the full 
wavelength range between 43 and 197\,$\mu$m with a spectral resolving 
power of $\sim\,$200 and can be utilised to search for spectral 
signatures  related to the asteroids' mineralogic composition. 
An overview of the spectroscopic features from theory and lab measurements
in the infrared wavelength regime up to 25\,$\mu$m can be found in e.g. \citet{Salisbury:1993} 
and in the context of ISO in \citet{Dotto:2002}. 
Far infrared laboratory spectra of various crystalline and amorphous minerals are digitally
available e.g. in the Heidelberg -- Jena -- St.Petersburg -- Database of Optical Constants (HJPDOC)\footnote{\tt http://www.mpia-hd.mpg.de/HJPDOC/} \citep{Henning:1999}.
Although these measurements are in principle sufficient to allow the analysis 
of the data presented in this paper, their application to solid surfaces with non-uniform
grain sizes and subsurface radiation is not straightforward, rendering a mineralogic
interpretation difficult.

Apart from that, the LWS observations can  give valuable input to models describing the thermal 
emission of minor bodies in our own or other solar systems, and are key 
ingredients to establish asteroids as calibration objects in the middle and 
far infrared \citep[e.g.][]{Muller:1998,Muller:2002}.

An effort was made to produce a homogeneously reduced set of these
observations, eventually leading to a catalogue which is available
through the IDA. 
We describe the contents of the catalogue (Section~\ref{sec:obs}), outline the
data reduction process, and give an overview of the instrumental artefacts 
(Section~\ref{sec:reduction}) encountered during catalogue compilation. 
Section~\ref{sec:catalogue} gives a brief overview of the catalogue structure. 
In Section~\ref{sec:quality} we discuss the quality of the final products, present  
diameter and albedo values derived from thermal IR data in the range 43--197$\mu$m
for the observed asteroids, and analyse the thermal 
lightcurve of (4)~Vesta.


\section{Observations}
\label{sec:obs}

\subsection{Overview}
  
Most of the LWS asteroid observations were carried out in fixed grating mode,
using the astronomical observing template (AOT) L02. 
This produced ten photometric measurements -- one in each detector -- at 46.2\,$\mu$m, 
56.2\,$\mu$m, 66.1\,$\mu$m, 75.7\,$\mu$m, 84.8\,$\mu$m, 102.4\,$\mu$m, 141.8\,$\mu$m,
160.6\,$\mu$m, and 178.0\,$\mu$m (corresponding to detectors SW1 to SW5
and LW1 to LW5). 

Only four full grating scans using the template L01 were performed, 
covering the whole spectral range from 43 to 197\,$\mu$m with typical SNRs
of 20--30 for (10) Hygiea and 150--200 for (1) Ceres.
The spectrum is composed of ten sub-spectra with the sub-spectra being
generated by the grating scanning over the ten LWS detectors simultaneously.
Signal-to-noise was built up by taking more than one scan.

Additionally there exist 25 observations carried out in the non-standard 
engineering mode L99, most of them similar to L01 grating scans 
(SNR $\approx$150 for (1) Ceres and $\approx$100 for (4) Vesta).
For the catalogue, only L99 on-source grating scans were considered, and
no background measurements are included.
The non-standard observations are characterised by experimental settings
of the detector parameters (bias voltages and heater currents), an unusually
high spectral sampling (13 samples per resolution element instead of four,
as it was the standard setting later), and longer integration times per grating
position.

From the available L99 observations we selected seven datasets for inclusion in
this catalogue. 
The spectral analysis of this data still remains difficult, since the relative 
spectral response function (RSRF) of the detectors used in the calibration 
process is based on observations performed in the standard L01 mode.
The remaining L99 data sets did not include sufficient calibration data
for reliable dark current determination.

All standard observations are listed in Tbl.~\ref{tbl:standard}, while the
non-standard observations are summarised in Tbl.~\ref{tbl:nonstd}. 
Contamination by 158\,$\mu$m CII background emission is 
indicated where applicable.

A comprehensive overview of the instrument design, observing modes, and 
detector properties can be found in the LWS handbook \citep{Gry:2003}.

\addtocounter{table}{1}
\addtocounter{table}{1}


\subsection{Observation elements}

Each observation consists not only of flux measurements of the target itself, 
but also includes calibration data, typically split in two blocks at begin 
and end.
By placing the Fabry-Per\'ot spectrometer into the beam and intentionally 
misaligning the etalons, light from the source did not reach the detectors, 
eventually resulting in a dark current measurement. 
Long observations, and especially the engineering mode L99 data, contain
intermittent additional dark current measurements.

Since the responsivity of the detectors is not constant, but
usually drifting upwards over time due to impacts of charged
particles, the actual responsivity of each detector had to
be determined for each observation separately.
This was done by flashing the detectors with internal illuminators 
while light from the source was again blocked with the Fabry-Per\'ot.
This allowed to quantify responsivity changes during the course 
of a single observation, and to apply the so-called 'responsivity 
drift correction'.

Absolute responsivity calibration was
performed by comparing the responsivities measured at the time
of the observation to the ones measured during observations
of \object{Uranus}, the primary LWS calibrator.

For the target source flux measurements, the Fabry-Per\'ot was removed
from the beam. 
Depending on the observation template and requested
wavelength coverage, the grating remained either at a fixed angle or
was moved to scan the desired spectral range. 
This could be done in unidirectional mode, where the grating angle 
constantly in- or decreased and was reset to its start position before each 
scan repetition. 
Alternatively, the observation could be performed in bidirectional mode, 
with the grating angle increasing and then decreasing with the same rate 
during the measurement. 
This is referred to as up- and down-scan, where 'up' and 'down' does not 
correspond to an  increase or decrease  of the observed wavelength, but 
of the grating position encoder value. In fact, during an upscan the observing 
wavelength decreased, and increased during downscans.


\section{Data reduction}
\label{sec:reduction}

\subsection{Instrumental artefacts}
\label{subsec:artifacts}

The most prominent artefacts, common to all ISO instruments,
are 'glitches', sudden increases of the measured flux due to
charged particles hitting a detector or parts of the readout
electronics. The initial sudden signal increase is sometimes
followed by a `glitch tail', a long-lived exponential decay.
Though glitches can be recognised and filtered out automatically 
to some extent, glitch remnants are present in all observations 
throughout the archive, regardless of instrument or observing mode.
Glitches do not only occur during object flux measurements, but
also during the calibration, i.e. dark current and responsivity
measurements, and adversely affect the accuracy and reliability
of the actual observation. 
Special attention was therefore put on the manual removal of glitch 
remnants in each observation contained in the presented catalogue.

Other frequently observed instrumental artefacts are fringes in
the spectra in the case of extended sources or off-axis observations,
spurious spectral features due to uncertainties in the spectral
response function used in the calibration process, and memory
effects, i.e. changes of detector response depending on the
illumination history.

Observations towards the end of the satellite mission were
affected by so-called 'warm-up effects', visible as broad
spectral features in some detectors. 
This was presumably caused by periodic breaks of the liquid 
helium film in the vicinity of the LWS strap location as the satellite's 
helium  tank came close to exhaustion.
From the observations contained in this catalogue, this affects one L01 type
observation of (1) Ceres taken in December 1997, several months before ISO's 
liquid helium depletion in April 1998.

A thorough description of caveats and unexpected effects concerning
LWS data can be found in Chapter 6 of the LWS handbook \citep{Gry:2003}.
All datasets included in this catalogue were checked for the presence
of the artefacts described above and data quality was quantified via
a set of well-defined flags. 
The criteria for setting these flags and the
results of this quality check can be found in the technical documentation
accompanying the catalogue \citep{Hormuth:2006}.


\subsection{Data reduction steps}

Data reduction was performed with the available standard reduction packages 
LIA (LWS Interactive Analysis) Version 10.2 and ISAP (ISO Spectral Analysis Package) 
Version 2.2. 
Additional own IDL-procedures were used for manual editing of the data at the 
SPD (Standard Processed Data) level, and the generation of the final FITS 
and ASCII files. The SPD data is preprocessed in the sense that raw detector readouts have been
converted into photocurrents, saturated readouts flagged, and strong glitches
automatically identified and removed. Housekeeping data not necessary for further data reduction
has been removed, but dark current and responsivity measurements have been preserved.
In the following we briefly outline the order of reduction
steps and give the names of the corresponding LIA procedures where applicable.

In a first step, the pointing coordinates of each observation were checked against the
ephemeris of the observed asteroid to detect pointing problems. 
The maximum difference between asteroid position and telescope pointing was 
found to be 2\farcs5, much smaller than the LWS  beam profile
 \citep[see][]{Lloyd:2003}.
The reduction process started at SPD level with a visual inspection of 
the data in the time domain. 
Photometric L02 observations were manually de-glitched at this stage, 
capturing minor glitches not automatically flagged during generation of the
SPD data.
The available pipeline data of grating scans was additionally inspected with 
ISAP to get a quick  overview of the available scans and scan directions. 

Next, dark current determination and subtraction were performed using the 
LIA routine \texttt{IA\_DARK}, followed by the absolute responsivity 
correction with \texttt{IA\_ABSCORR}. 
These interactive routines allow visual inspection and manual editing of the dark current measurements
and the internal calibration source observations. Deglitching of this data and the
removal of measurements obviously affected by memory effects is critical, since any
artefacts in the calibration data will affect the accuracy and reliability of the
whole observation.
The LIA routine \texttt{SHORT\_AAL} was used to create ISAP-compatible
output spectra in the LWS Auto-Analysis format, and to convert to 
W/cm$^2$/$\mu$m, the flux unit used for all LWS observations. This routine automatically
strips all calibration data and leaves only the basic information needed for
further scientific analysis, e.g. wavelength, flux, detector number, and scan direction.

In the case of photometric L02 observations, the data was now 
averaged per detector to create the final FITS product.

In the case of grating scans a first round of manual de-glitching was 
performed, probably the most time consuming step in the reduction process.  
Relative responsivity correction was achieved by comparing the mean
flux in each detector and scan to the mean flux level of all scans together.
The derived gains were then used to bring all scans  to the same
flux level -- separately for each detector and scan direction.

After fine-zapping of remaining glitches, the data was averaged per detector 
with a bin width  of typically 0.06\,$\mu$m, corresponding to an oversampling 
of four with respect to the spectral resolution. 
In the case of bidirectional scans the data was now averaged over the 
scan directions. 
Especially at longer wavelengths some spectra clearly showed fringing,
probably caused by background emission. 
The affected detectors were defringed within ISAP, using only the averaged 
spectrum for fringe detection and defringing to get the best possible signal-to-noise 
for the fringe fitting (in the case of bidirectional data).

   \begin{figure}[h!tb]
   \centering
     \includegraphics[width=6.5cm,angle=90]{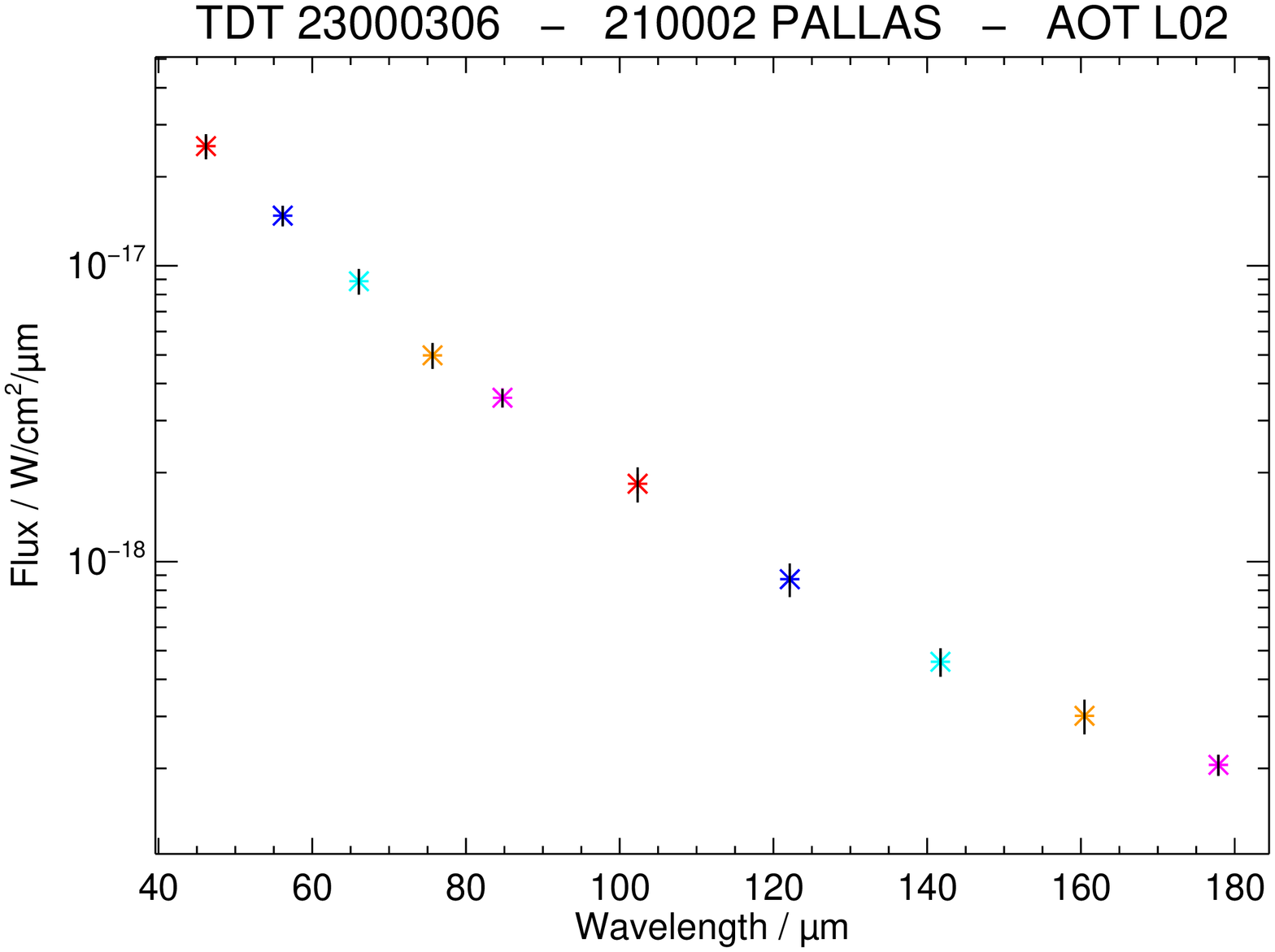}
     \includegraphics[width=6.5cm,angle=90]{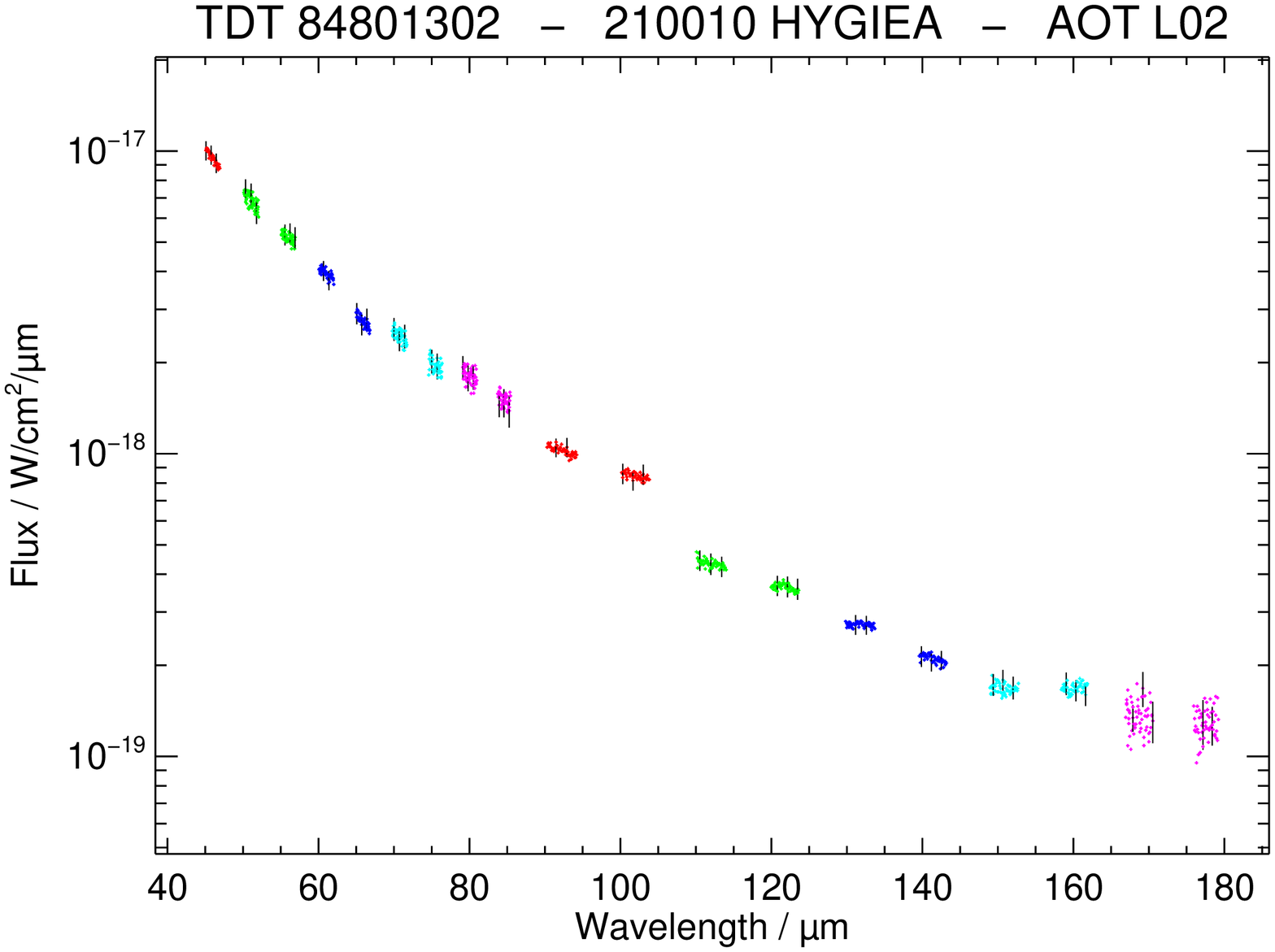}
     \caption{Examples for L02 observations. Left: fixed grating observation of (2)~Pallas.
                     Right: short grating scan observation of (10)~Hygiea.}
              \label{fig:datal02}
    \end{figure}
   \begin{figure}[h!tb]
   \centering
     \includegraphics[width=6.5cm,angle=90]{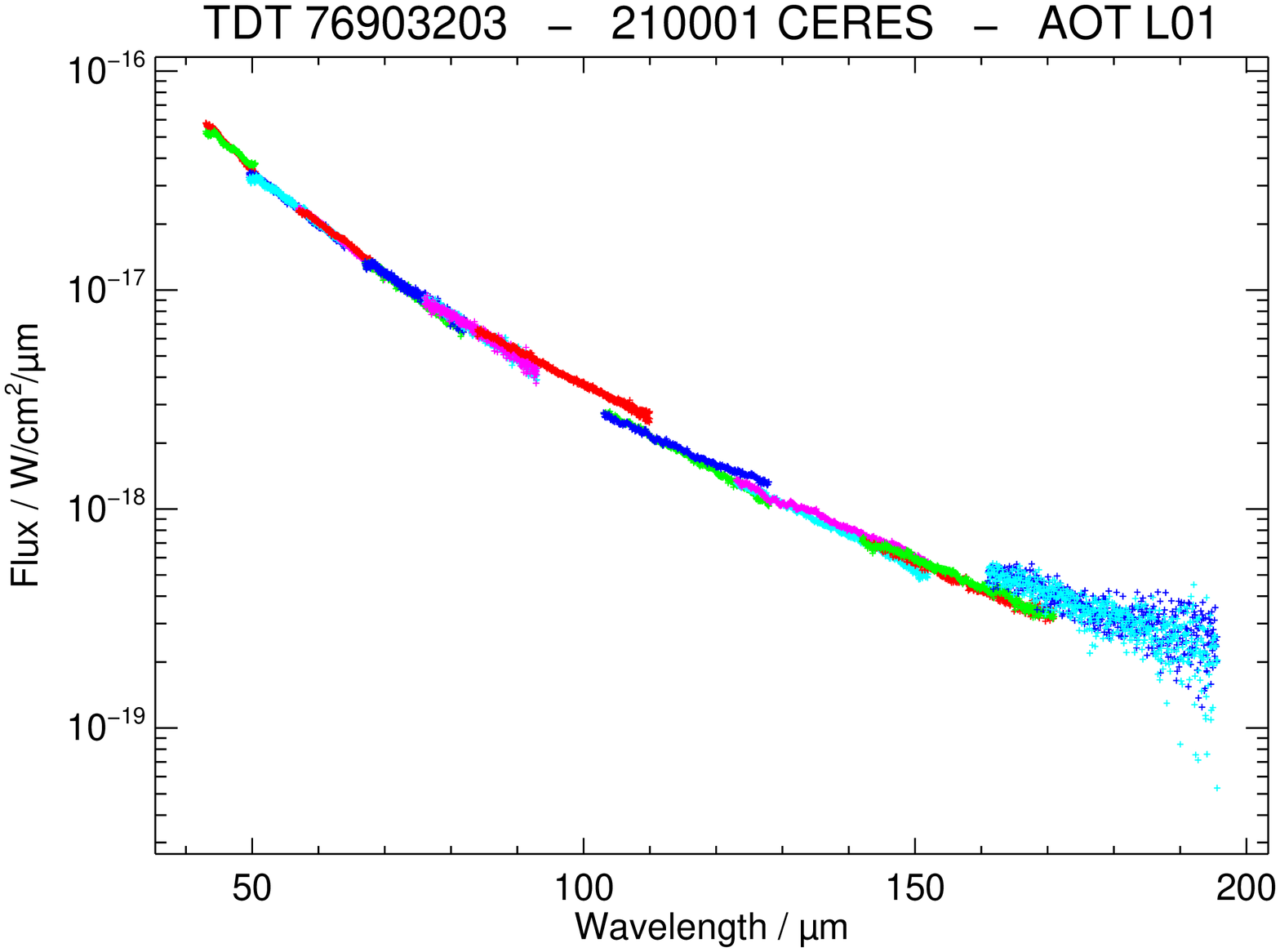}
     \includegraphics[width=6.5cm,angle=90]{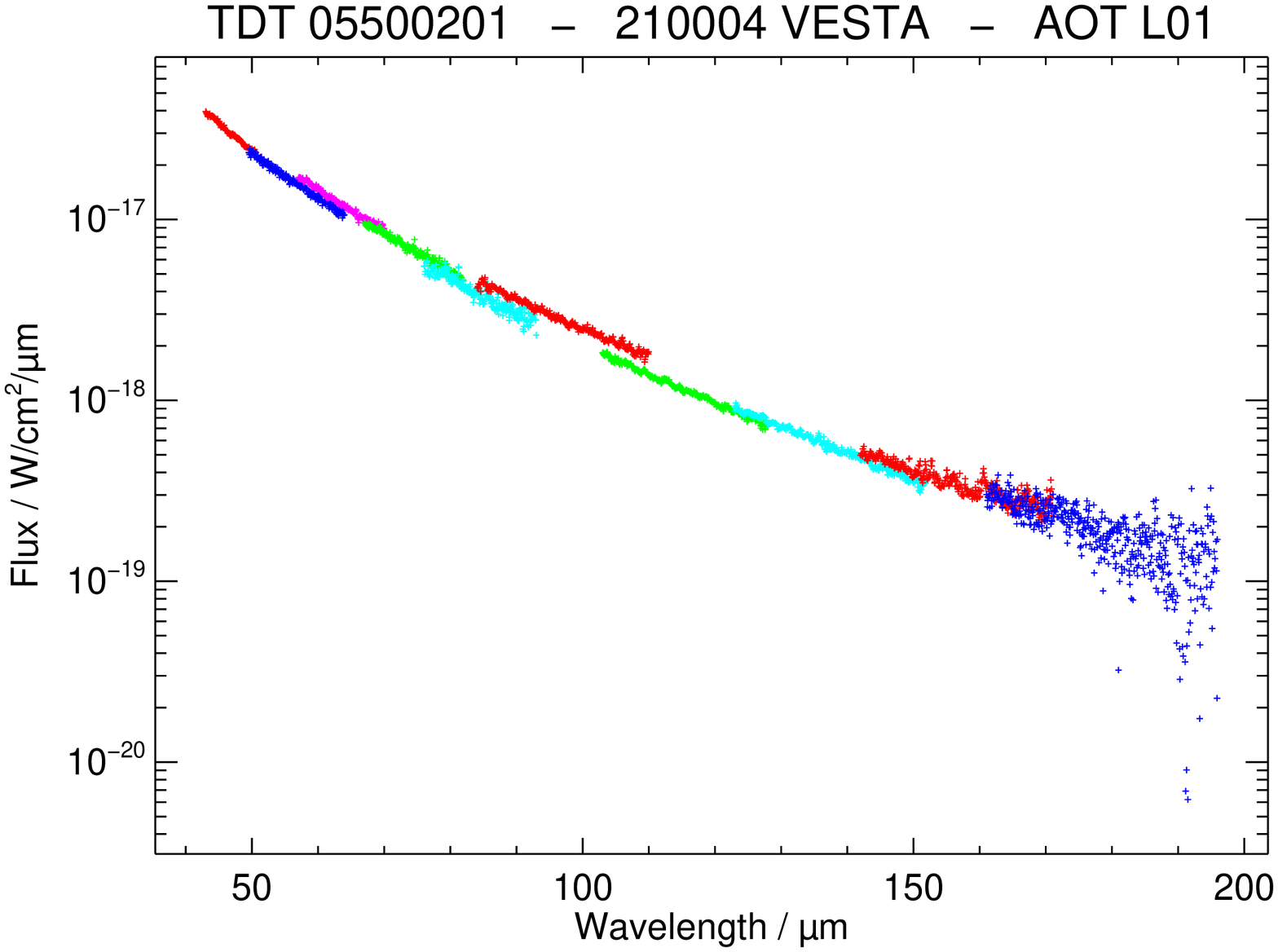}
     \caption{Examples for full wavelength range scans. Left: standard L01 observation of (1)~Ceres.
                     Right: Engineering mode L99 observation of (4)~Vesta.}
              \label{fig:datal01}
    \end{figure}

\section{Data products and access to the catalogue}
\label{sec:catalogue}

The ISO Data Archive (IDA)\footnote{Accessible via \texttt{http://iso.esac.esa.int}} contains
spectroscopic, imaging, photometric, and polarimetric measurements of more than 40 
different asteroids at wavelengths between 2 and 240$\mu$m. A versatile Java based
web-interface allows to extract observational data in various formats, ranging from unprocessed
raw data, over standard pipeline processed measurements up to manually reprocessed
high quality data products.

All data files associated to this catalogue can be retrieved by searching for the specific
TDT\footnote{``Target Dedicated Time'', a unique identifier for ISO observations} number of a measurement as given in Tbl.~\ref{tbl:standard} and \ref{tbl:nonstd},
by searching for all LWS observations of a given asteroid, or by searching for LWS
asteroid observations in general. The highly processed
data products, including overview plots, quality flags, and catalogue documentation
can be obtained via the HPDP button left to each entry in the search result window of 
the archive interface. \\
The catalogue documentation \citep{Hormuth:2006}, giving a thorough description of the data reduction steps,  
is also separately available from
the list of all HPDP data sets\footnote{\texttt{http://iso.esac.esa.int/ida/hpdp.php}}.
      
In Fig.~\ref{fig:datal02} and Fig.~\ref{fig:datal01} we show logarithmic example plots  for all observing modes. 
Different colours have been
chosen to allow distinguishment of overlapping measurements by different
detectors.
The error bars in Fig.~\ref{fig:datal02} include measurement and calibration errors,
  incuding a 5\% uncertainty attributed to the Uranus models used for
  absolute flux calibration.
The slight mismatches between measurements from different
  detectors at same wavelengths in Fig.~\ref{fig:datal01} are caused by memory effects
  of individual detectors.

The data sets are available both as FITS tables, compatible with the ISAP
software package, and as plain ASCII tables. Please refer to the catalogue
documentation for information about individual FITS header keywords and table columns.


\section{Discussion}
\label{sec:quality} 

\subsection{Quality of the catalogue products}
In order to study the reliability of the photometric L02 data in the
catalogue, we compared the measured flux with predictions calculated
by means of the thermophysical model \citep[TPM; e.g. ][and references therein]{Lagerros:1998}.   
For Tbl.~\ref{tbl:l02-err} we calculated model
fluxes for the specified targets and epochs and analysed the observation-to-model
ratios. Measurements affected by detector warm-up artefacts or abnormally high
dark current were not included in this quality check.

  \begin{table}[h!tb]

  \caption{Ratios between LWS\,02 observations and TPM predictions,
           weighted mean values and standard deviations per detector
           (SW1, 2, 3, 4, 5 and LW1, 2, 3)
           and per asteroid. The photometry of the LW4 and LW5 detectors
           is less reliable due to dark current problems.
       \label{tbl:l02-err}}
\centering
  \begin{tabular}{rcccc}
  \noalign{\smallskip}
  \hline\hline
  \noalign{\smallskip}
                      & (1) Ceres      & (2) Pallas     & (4) Vesta      & (10) Hygiea \\
\mbox{$\lambda$}   & 11 Obs.    & 4 Obs.     & 14 Obs.    & 2 Obs. \\
\mbox{[$\mu$m]}    & Obs/Mod    & Obs/Mod    & Obs/Mod    & Obs/Mod \\
  \noalign{\smallskip}
  \hline
  \noalign{\smallskip}
46.4 & 1.04$\pm$0.05 & 1.00$\pm$0.02 & 1.02$\pm$0.04 & 1.21$\pm$0.05 \\
56.4 & 1.01$\pm$0.03 & 1.06$\pm$0.02 & 1.01$\pm$0.04 & 1.25$\pm$0.10 \\
66.3 & 1.04$\pm$0.02 & 1.10$\pm$0.04 & 1.02$\pm$0.04 & 1.23$\pm$0.15 \\
75.9 & 1.02$\pm$0.03 & 1.02$\pm$0.03 & 1.04$\pm$0.04 & 1.24$\pm$0.12 \\
 85.0 & 1.01$\pm$0.04 & 1.02$\pm$0.11 & 1.00$\pm$0.07 & 1.39$\pm$0.10 \\
  102.9 & 1.09$\pm$0.04 & 1.13$\pm$0.04 & 1.24$\pm$0.06 & 1.40$\pm$0.26 \\
 122.7 & 0.96$\pm$0.05 & 0.95$\pm$0.03 & 1.07$\pm$0.05 & 1.30$\pm$0.07 \\
142.2 & 0.89$\pm$0.04 & 0.84$\pm$0.07 & 1.10$\pm$0.11 & 1.33$\pm$0.04 \\
 161.0 & 0.99$\pm$0.05 & 0.99$\pm$0.04 & ---           & --- \\
 178.4 & 1.01$\pm$0.10 & ---           & ---           & --- \\
  \noalign{\smallskip}
  \hline
  \noalign{\smallskip}
  \multicolumn{1}{c}{Mean}
      & 1.01$\pm$0.05 & 1.03$\pm$0.08 & 1.05$\pm$0.08 & 1.27$\pm$0.14 \\
  \noalign{\smallskip}
  \hline
  \noalign{\smallskip}
  \end{tabular}
  \end{table}

  The general agreement between the photometric observations (L02)
  and models (Tbl.~\ref{tbl:l02-err}) is better than 5\,\% for
  Ceres, Pallas, and Vesta and better
  than 30\,\% for Hygiea. The higher uncertainties are
  due to dark current effects at the low flux levels. The discrepancy
  for Hygiea is not surprising. Due to the lack of suitable good
  quality thermal data the thermophysical model parameters are also
  very poor, and Hygiea is considered as a low quality far-IR calibrator \citep{Muller:1998}.
   As a rule of thumb, the absolute flux error is better than 10\% for
  targets brighter than about 10\,Jy.

  The high quality of the spectral scans can be seen in the 
  smooth and well-connected curves of the archive browse images
  (see Fig.~\ref{fig:datal01}).  For mineralogic studies it might still
  be useful to investigate the up- and down-scans separately
  to confirm that certain low-level, broad-band structures in
  the scans are reliable or not. Low level features extending over two
  or more detectors as well as structures close to pronounced RSRF
  changes have to be analysed carefully: the detectors are not
  always behaving in exactly the same way and LWS should be considered
  as 10 individual spectrometers, one per detector.


\subsection{Diameter and albedo calculations}

 \begin{table*}
  \begin{center}
    \caption{Object specific TPM input parameters. The $\lambda_{ecl}$
             and $\beta_{ecl}$ values are the ecliptic coordinates of
             the spin vector direction, with $\beta_{ecl}$ counted
             from the equator.
         \label{tbl:tpm_input}}
  \begin{tabular}{ccrcl}
  \noalign{\smallskip} \hline \noalign{\smallskip}
  Object      & [H,G]              & $P_{sid}$ [h] & $\lambda_{ecl}$, $\beta_{ecl}$ & source of shape \& spin vector\\
  \noalign{\smallskip} \hline \noalign{\smallskip}
   (1) Ceres  & [3.28,0.05]$^{1}$  &  9.074184 & 331.51$^{\circ}$, +77.88$^{\circ}$ & HST observations$^{4}$ \\
   (2) Pallas & [4.13,0.16]$^{2}$  &  7.813225 &  34.80$^{\circ}$, -11.75$^{\circ}$ & Lightcurve inversion techniques$^{5,6}$ \\
   (4) Vesta  & [3.20,0.34]$^{3}$  &  5.342129 & 319.45$^{\circ}$, +59.23$^{\circ}$ & HST observations$^{7}$ \\
  (10) Hygiea & [5.43,0.07]$^{2}$  & 27.623265 & 117.01$^{\circ}$, -28.53$^{\circ}$ & Lightcurve inversion techniques$^{8}$ \\
  \noalign{\smallskip} \hline \noalign{\smallskip}
  \end{tabular}
  \end{center}
  $^{1}$ \citet{Lagerkvist:1992}; 
  $^{2}$ \citet{Lagerkvist:2001}; 
  $^{3}$ \citet{Muller:1998}; 
  $^{4}$ \citet{Thomas:2005}; 
  $^{5}$ M.\ Kaasalainen, priv.\ comm.\ 22/Sep/2005;
  $^{6}$ \citet{Torppa:2003}; 
  $^{7}$ \citet{Thomas:1997}; 
  $^{8}$ M.\ Kaasalainen, priv.\ comm.\ 31/Jan/2006;
  \end{table*}

Direct size measurements, even for the large main-belt asteroids, are
still difficult. Up to now, only a small sample has been resolved via
imaging techniques from space (HST, see \citet{Dotto:2002}
and references therein) and adaptive optics techniques from large ground-based
telescopes \citep[e.g.][]{Marchis:2006}. Occultation observations
performed by well-organised groups of amateur astronomers also reveal direct
sizes, expressed in chord maps which represent the cross section of the
asteroid at the time of the occultation \citep[e.g.][]{Millis:1989}
or http://www.psi.edu/pds/resource/occ.html for more recent results).
But the largest number of asteroid sizes are originating from thermal
infrared observations \citep[e.g.][]{Tedesco:2002a,Tedesco:2002} via radiometric methods i.e., 
an indirect method which relies on model assumptions.

We used a well-established thermophysical model \citep[TPM;][]{Lagerros:1996,Lagerros:1997,Lagerros:1998} to derive absolute effective sizes
and albedos for our 4 asteroids from the new L02 data sets.

This model allows to use state-of-the-art shape models together with
corresponding spin axis solutions derived from lightcurve inversion techniques
\citep[e.g.][]{Kaasalainen:2002}. As part of the energy equation for the asteroids the
amount of reflected sunlight is described via the H$-$G magnitude system \citep{Bowell:1989}.
The rest of the solar energy is absorbed and re-emitted as thermal emission. Knowing the solar
insolation together with the disk-integrated thermal emission allows then to solve the energy
equation for the asteroid's effective size and albedo \citep[e.g.][]{Harris:2002}.
The true illumination and observing geometries are taken into account in the TPM
calculations. The surface roughness is described by the r.m.s. values of the surface slopes ($\rho$)
and the fraction of the surface covered by craters (f). For the surface roughness, as well as for
the thermal properties to calculate the heat conduction into lower layers of the surface, we used
``default properties'' ($\rho$=0.7, f=0.6, $\Gamma$=15\,J\,m$^{-2}$\,s$^{-0.5}$\,K$^{-1}$)
as described by \citet{Muller:1999}.
The emissivity models were taken from \citet{Muller:1998}: a standard wavelength-dependent
emissivity model for Ceres, Pallas, ad Hygiea and a special one for Vesta where lower 
emissivities in the submm/mm range were found. \\
The TPM with the ``default thermal properties'' has been tested and validated extensively
in the context of ISO for large, regolith-covered main-belt-asteroids \citep{Muller:2002}.
It is worth to note here that the surface roughness and heat conduction properties are less
important for our far-IR analysis since they mainly influence the mid-IR radiation at
the Wien-part of the spectral energy distribution \citep[e.g.][Figs. 2\&3]{Muller:2002b}. \\
Asteroid-specific model input parameters and the corresponding references are listed
in Tbl.~\ref{tbl:tpm_input}.

Our derived diameter and albedo values are shown in Tbl.~\ref{tbl:tpm_results}.
The diameters are effective diameters of a sphere of equal volume. It can
be considered as a ``scaling factor'' for shape models which lack the absolute
size information.
These values are the weighted mean values, based on the photometrically more reliable
short wavelength detectors (46.4 to 85.0\,$\mu$m). We also excluded
the data which are flagged as being affected by high dark currents or high glitch
rates. In the table we give the full standard deviation of all N radiometric
solutions to account for the fact that different observations see the asteroid
under different observing and illumination geometries.

\begin{table}[h!tb]
      \caption[]{Radiometric diameter and albedo values, derived
                 from the LWS L02 measurements.}
         \label{tbl:tpm_results}
         \centering
\begin{tabular}{lccc}
\hline
\hline
\noalign{\smallskip}
Object & D$_{eff}$ [km] & p$_{V}$  & N \\
\noalign{\smallskip}
\hline
\noalign{\smallskip}
 (1) Ceres  &  959.6 $\pm$ 16.9 & 0.096 $\pm$ 0.003 & 55 \\
 (2) Pallas &  534.4 $\pm$ 14.7 & 0.142 $\pm$ 0.009 & 20 \\
 (4) Vesta  &  548.5 $\pm$ 12.5 & 0.317 $\pm$ 0.015 & 70 \\
(10) Hygiea &  469.3 $\pm$ 26.5 & 0.056 $\pm$ 0.006 & 20 \\
\hline
\end{tabular}
\end{table}

These diameter and albedo values are the first ones which are
derived purely from far-IR observations. Nevertheless, the
results agree very well with the so far published values.

he most accurate values for the diameter of (1)~Ceres come
 from HST observations \citep{Thomas:2005} and adaptive optics (AO)
 observations using the Keck II telescope \citep{Carry:2008}.
 The HST observations led to an oblate spheroid with axes of
 $a$=$b$=487.3\,km and $c$=454.7\,km, resulting in an effective
 diameter of $2\times(a \times b \times c)^{1/3} = $952.4\,km.
 The AO data confirmed the oblate spheroid, but with slightly
 different values (a = b = 479.7$\pm$2.3\,km, c = 444.4$\pm$2.1\,km)
 and an effective diameter of 935.3\,km. 
 Our result is derived from very limited viewing geometries, but it is in excellent 
 agreement with the HST results. 
 In this context it is interesting to note that the diameter derived
 from IRAS observations at 12, 25, 60 and 100\,$\mu$m was 848.40$\pm$19.7\,km
 \citep{Tedesco:2002}. Also our albedo value agrees very well
 with the 0.0936 derived via stellar occultation techniques
 \citep{Shevchenko:2006}.
 
(2)~Pallas: \citet{Tedesco:2002} obtained from IRAS data values of
498.07$\pm$18.8\,km and a geometric albedo of 0.1587$\pm$0.013.
A combination of speckle, occultation and lightcurve observations
of (2)~Pallas lead to an ellipsoidal shape with dimensions of
$574 \pm 10 \times 526 \pm 3 \times 501 \pm 2$ km and a mean diameter
of 533 $\pm$ 6\,km \citep{Dunham:1990}. 
A recent high-angular resolution adaptive optics study by \citet{Carry:2007}
resulted in an absolute effective size of 519.5\,km (semi-major ellipsoidal axes
values of 276$\times$256$\times$248\,km, $\pm10$\,km). 
\citet{Drummond:2008} found 494$\pm$57\,km based on
 another set of AO observations. The various results show that there
 are still significant uncertainties in the overall shape and size
 values, but the derived effective sizes agree within the given
 errorbars with our radiometric diameter. Therefore, we believe that
 our albedo of 0.142 is the most reliable value so far published for Pallas.

(4)~Vesta has a very complex surface with large albedo variations
on the surface \citep{Binzel:1997}. Our diameter value is about 4\%
larger than the effective diameter from HST observations (axes 289, 280,
and 229 $\pm$ 5 km; D$_{eff}=$ 529.2\,km; \citet{Thomas:1997}, note added in proof)
and about 7\% larger than the mean triaxial ellipsoid
 solution with $563 \pm 5 \times 534 \pm 5 \times 442 \pm 7$\,km
 ($2\cdot(a\cdot b\cdot c)^{1/3} = 510.3 \pm 5.6\,km$) given by
\citet{Drummond:2008}. 
This is outside the given rms-scatter from our 70 measurements which
might be an indication that the special emissivity model for Vesta
\citep{Muller:1998} needs a small adjustment for this wavelength
range. 
It looks like the emissivity drop which has been seen by \citet{Redman:1992,Redman:1998}
occurs already at shorter wavelengths below 50\,$\mu$m.
Almost all LWS observations of Vesta were taken on one day (revolution 805)
under very similar aspect angles. 
This might also influence the outcome
of the TPM technique which works best when combining data from
different wavelengths, phase angles, rotational phases and aspect angles
\citep{Muller:2002}. 
\citet{Shevchenko:2006} derived an albedo of p$_{H}$=0.370
from occultation measurements,  while we obtained a radiometric value of p$_{V}$ =
0.32$\pm$0.02, again this discrepancy might be explained by emissivity
effects or the limited aspect angle range.
The albedo value in Tbl.~\ref{tbl:tpm_results} agrees 
very well with previous studies by \citet{Muller:1998} (p$_{V}$ = 0.33)
based on a set of thermal observations from mid-IR to mm-wavelengths.

Our result for (10)~Hygiea deviates significantly from previous studies: 
\citet{Tedesco:2002} found values of 407.12$\pm$6.8\,km and p$_{V}$=0.0717$\pm$0.002. 
\citet{Muller:1998} gave 429.9\,km and 0.066. 
But Hygiea's shape model is not that well defined due to a poor lightcurve
coverage (in rotational phases) and some very low quality lightcurves (Kaasalainen, priv.\ comm.).
There might also be a second pole solution at around $\lambda_{ecl}$ = 300\degr\ (instead of the
117\degr\ used here). But neither using the second pole solutions nor using a spherical
shape model lower the predicted size values (or the standard deviations) significantly.
One possibility for the discrepancies in the diameters could be emissivity issues.
If Hygiea's far-IR emissivity is lower then the corresponding diameter would also be
smaller (and the albedo a bit larger). But the data quality and the very limited aspect
angle range (all LWS observations were taken within 20\,days) are not sufficient 
to draw firm conclusions.
Especially when looking at the best quality direct size measurement by \citet{Ragazzoni:2000}:
They gave an effective diameter of 444$\pm$35\,km (with an axis-ratio of a/b=1.11),
based on speckle techniques. Our radiometric results are still well within these error bars.


\subsection{Far-IR thermal lightcurve of (4)~Vesta in the range 45$-$105\,$\mu$m}

The visual lightcurve of (4)~Vesta
is dominated by the influence of the albedo variations
 \citep{Degewij:1979}.
Standard lightcurve inversion techniques failed to produce
a reliable shape model (Kaasalainen, priv. comm.), but high
resolution HST imaging allowed a solution for the shape and spin
vector \citep{Thomas:1997}.
\citet{Redman:1992} found that the 1\,mm light-curve
is apparently dominated by the triaxial shape, without any significant contributions
from the optical albedo spots. 
\citet{Muller:2007} showed that at even longer wavelengths (around 3.2\,mm) the
mm-lightcurve follows for a large fraction of the
rotational period the shape-introduced variations. They also demonstrated that the rotational
phases with clear deviations are connected to structures (e.g., the Olbers feature)
which are visible in the HST images of (4)~Vesta \citep{Binzel:1997}. 

Before we combined TPM lightcurve predictions with the observed fluxes, we explored
the influence of TPM input parameters on the thermal lightcurve amplitude and phase
for the given wavelength range and aspect angles. It turned out that neither the thermal
inertia (the range between 0 and 50\,J\,m$^{-2}$\,s$^{-0.5}$\,K$^{-1}$ was considered)
nor the beaming parameter, describing various roughness scenarios, influenced 
the predictions for our case significantly. The predicted lightcurve amplitudes
varied between 5$-$6\% (peak-to-peak) and the lightcurve phases only by a few
minutes between the low and high thermal inertia predictions.
Our data sets were therefore not sufficient to confine any of these parameters. 

The Vesta L02 observations from revolution 805 (28-Jan-1998) were taken over a period
of about 6\,hours, covering slightly more than one full rotation period of 5.34\,hours.
On the basis of the shape and spin-vector solution given in Tbl.~\ref{tbl:tpm_input}
and the diameter and albedo results from Tbl.~\ref{tbl:tpm_results} we predicted the
thermal lightcurves at the L02 key-wavelengths for the LWS observations
(note that the reference time frame is that of the ISO satellite and not the
asteroid-intrinsic one).

\begin{figure}[h!tb]
    \begin{center}
    \includegraphics[angle=90,width=9.3cm,trim=0cm 0cm 0cm 1.5cm]{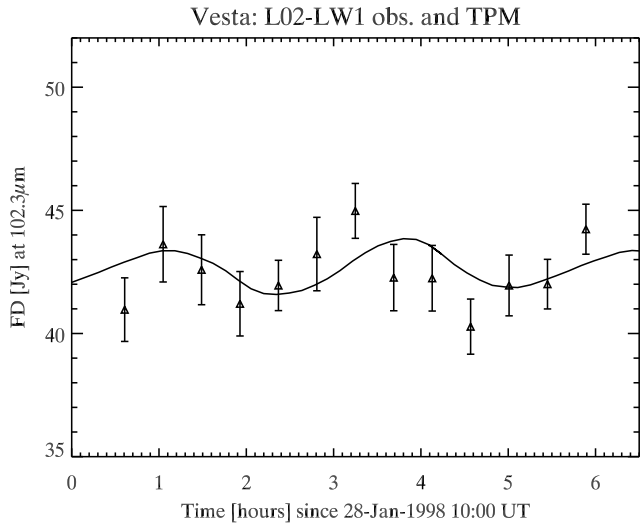}
    \caption[]{The observed (dots with error bars) and the predicted thermal lightcurve 
    	(continuous line) of (4)~Vesta
               at 102.3\,$\mu$m (detector LW1). }
    \label{fig:lc_lw01}
    \end{center}
\end{figure}

\begin{figure}[h!tb]
    \begin{center}
    \includegraphics[angle=90,width=9.3cm,trim=0cm 0cm 0cm 1.5cm]{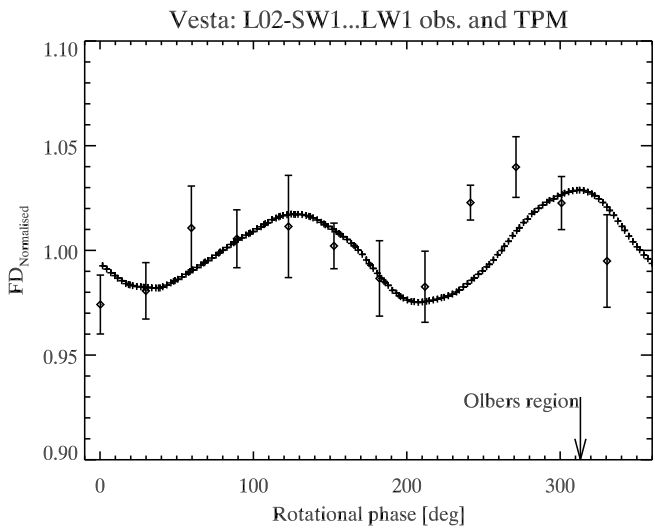}
    \caption[]{The observed (dots with error bars, normalised and averaged over five detectors (SW1, SW2, SW3, SW4, SW5 and
               LW1) and the predicted thermal lightcurve (continuous line) of (4)~Vesta as a function of
	       the rotational phase.}
    \label{fig:lc_obsmod}
    \end{center}
\end{figure}

Figure~\ref{fig:lc_lw01} shows the observations at 102.3\,$\mu$m together with the model
prediction on an absolute flux and time scale. Figure~\ref{fig:lc_obsmod} shows the
combined data below 105\,$\mu$m, first normalised per detector over the full observing
period and then averaged for a given rotational phase angle. The quality of the
data from the longer wavelength channels beyond 105\,$\mu$m was not sufficient
for this kind of analysis.

Both figures show that the observations  follow the model predictions. The
errorbars in Fig.~\ref{fig:lc_obsmod} correspond to the error of the mean values
plus a contribution from the flux uncertainty in these detectors. This figure combines
in total 78 independent measurements (6 detectors $\times$ 13 measurements at different
rotational phases). The model predictions match the observed lightcurve amplitude
as well as the lightcurve phase (in the observer's time frame). We also indicated the
rotational phase where the Olbers-feature \citep{Zellner:1997,Binzel:1997}
on Vesta's surface would be visible. We cannot see any influence of this surface feature
with lower albedo. \citet{Muller:2007} presented a more detailed discussion
on how various thermal properties of this surface feature would influence the thermal
lightcurve variations at different wavelengths.
They found strong emissivity variations which they attributed to the Olbers
structure and a neighbouring region. The deviation from the model predictions at
rotational phases between 240$-$280\degr\ in Fig.~\ref{fig:lc_obsmod} are also close 
to the Olbers region and might therefore be the far-IR signature of the ejecta 
material which is deposited on one side of the Olbers impact structure
as discussed in \citet{Muller:2007}.

\section{Conclusions}
Via radiometric techniques we derived the first far-IR-based
diameter and albedo values for the four large asteroids Ceres, Pallas, Vesta,
and Hygiea. Based on photometry in the wavelength range 46--85$\mu$m we
calculated effective diameters $D_{eff}$ of 960$\pm$17\,km, 534$\pm$15\,km, 549$\pm$13\,km,
and 469$\pm$27\,km for these four objects. The corresponding albedo values 
$p_V$ are 0.096$\pm$0.003, 0.142$\pm$0.009, 0.317$\pm$0.015, and 0.056$\pm$0.006, respectively. 
We found
excellent agreement with direct measurements from HST (Ceres
and Vesta) and speckle/occultation results (Pallas and Hygiea).
Our data contain also the first far-IR lightcurve of an asteroid:
Vesta's thermal lightcurve follows nicely the shape-introduced
variations. 
Albedo and grain effects are important in the visual and mm-lightcurves,
but  seem to play a very small role at far-IR wavelengths.

The catalogue of LWS observations of asteroids contains
 photometric and spectroscopic measurements of the main
 representatives of the main asteroid belt. The data will
 remain useful for thermophysical and mineralogical studies
 of these bodies, especially in the context of the {\em Dawn}
 space mission to Ceres and Vesta.



\begin{acknowledgements}
  We would like to thank Tim Grundy and his colleagues at RAL for preprocessing
  the non-standard data, and for the explanations concerning  the
  peculiarities of these observations.\\
  LIA is a joint development of the ISO-LWS Instrument Team at
  Rutherford Appleton Laboratories (RAL, UK - the PI Institute) and
  the Infrared Processing and Analysis Centre (IPAC/Caltech, USA).\\
  The ISO Spectral Analysis Package (ISAP) is a joint development by
  the LWS and SWS Instrument Teams and Data Centers. Contributing
  institutes are CESR, IAS, IPAC, MPE, RAL and SRON.
\end{acknowledgements}


\bibliographystyle{aa}
\bibliography{lwsaster}


\longtab{1}{
\begin{center}
\begin{longtable}{ll l l l l l l l}
      \caption{
               \label{tbl:standard} 
               Overview of standard LWS asteroid observations in our catalogue. For the on-source pointings the distance to
               the asteroid $\Delta$, the Sun-asteroid distance r, and the phase angle $\alpha$ are given, based on MPC
               (Minor Planet Center) ephemerides. The TDT number is a
               unique identifer, allowing to find the observation in the archive.}\\
    \hline\hline
    \noalign{\smallskip}
      \multicolumn{2}{l}{Object} & TDT &  {Mid-observation UTC}&  AOT &$\Delta$ $[$AU$]$ & r $[$AU$]$ & $\alpha$ $[\degr]$& {Remarks}\\
     \noalign{\smallskip}
    \hline
    \noalign{\smallskip}
    \endfirsthead
   \caption{continued}\\
    \hline\hline
     \multicolumn{2}{l}{Object} & TDT  & {Mid-observation UTC}&  AOT &$\Delta$ $[$AU$]$ & r $[$AU$]$ & $\alpha$ $[\degr]$& {Remarks}\\
    \hline
    \noalign{\smallskip}
   \endhead
   \noalign{\smallskip}
   \hline
   \endfoot
 (1) & Ceres   & 09300401 & 1996-02-18 \, 04:31:18 & L02 & 2.673 & 2.695 & 21.2 &  \\
     &         & 09304102 & 1996-02-18 \, 15:15:23 & L02 &  \multicolumn{4}{l}{(off-source reference measurement for TDT 09300401)} \\
     &         & 10500402 & 1996-03-01 \, 05:22:22& L02 & 2.520 & 2.705 & 21.5 &\\
     &         & 11900214 & 1996-03-15 \, 02:42:57 & L02 &2.344 &2.715&21.1&\\
     &         & 11905611 & 1996-03-15 \, 16:54:46 & L02 & \multicolumn{4}{l}{(off-source reference measurement for TDT 11900214)} \\   
     &         & 12600114 & 1996-03-22 \, 02:03:54 & L02 & 2.258&2.721&20.5& \\
     
     &         & 25800302 & 1996-07-31 \, 19:30:27 & L02 & 2.258 & 2.823& 19.2&\\
     &         & 25805903 & 1996-08-01 \, 08:55:54 & L02 &  \multicolumn{4}{l}{(off-source reference measurement for TDT 25800302)}\\ 
     
     &         & 26500301 & 1996-08-07 \, 19:10:27 & L02 &2.351 &2.828&20.0&\\
     &         & 26505602 & 1996-08-08 \, 09:32:38 & L02 &  \multicolumn{4}{l}{(off-source reference measurement for TDT 26500301)}\\
     
     &         & 32100204 & 1996-10-02 \, 15:31:35 & L02 & 3.126&2.868&18.6&\\
     &         & 32103506 & 1996-10-03 \, 03:52:44 & L02 & \multicolumn{4}{l}{(off-source reference measurement for TDT 32100204)} \\    
     
     &	       & 53802209 &  1997-05-07 \, 12:16:09 & L02 &3.112 &2.971&18.9&\\
     &         & 57902409 & 1997-06-17 \, 09:56:26& L02 & 2.575 &2.978&19.4&\\
     &         & 59401908 & 1997-07-02 \, 06:42:05 & L02 & 2.391&2.980&17.9& \\
     &         & 72001901 & 1997-11-05 \, 01:10:04 & L02 &2.522&2.973&18.6& \\
     
     &         & 74803304 & 1997-12-03 \, 03:04:58 & L02 &2.901&2.966&19.3& \\
     &         & 74803403 & 1997-12-03 \, 08:11:59 & L01 & 2.904&2.966&19.3&contains warmup artefacts \\     
     &         & 75502902 & 1997-12-10 \, 00:30:08 & L01 &  \multicolumn{4}{l}{(off-source reference measurement for TDTs 74803...)}\\
     
     &         & 75503003 & 1997-12-10 \, 01:38:00 & L02 &2.994&2.964&19.0& \\
     &         & 76200502 & 1997-12-16 \, 13:29:15 & L02 &3.081&2.962&18.6& \\
     &         & 76903102 & 1997-12-24 \, 01:08:50 & L02 &3.177&2.960&18.0& \\
     &         & 76903203 & 1997-12-24 \, 02:17:12 & L01 & 3.178 &2.960&18.0&\\

    \noalign{\smallskip}	
 (2) &	Pallas & 23000306 & 1996-07-03 \, 21:24:01 & L02 &2.497 &2.823&20.9&\\
     &         & 23002907 & 1996-07-04 \, 12:01:22 & L02 &\multicolumn{4}{l}{(off-source reference measurement for TDT 23000306)} \\
     &         & 25100202 & 1996-07-24 \, 19:49:12 & L02 &2.780&2.871&20.6& \\
     &         & 25103603 & 1996-07-25 \, 09:58:26 & L02 & \multicolumn{4}{l}{(off-source reference measurement for TDT 25100202)}  \\
     &         & 26500503 & 1996-08-07 \, 19:43:17 & L02 &2.967&2.903&19.9& \\
     &         & 26505204 & 1996-08-08 \, 07:27:57 & L02 &\multicolumn{4}{l}{(off-source reference measurement for TDT 26500503)}  \\
     &         & 27200203 & 1996-08-14 \, 18:19:41 & L02 &3.057&2.919&19.3& \\
     &         & 27202004 & 1996-08-15 \, 03:27:01 & L02 & \multicolumn{4}{l}{(off-source reference measurement for TDT 27200203)}  \\
	
     \noalign{\smallskip}
 (4) & Vesta   & 24402202 & 1996-07-17 \, 23:20:20 & L02 &1.593&2.149&26.6& \\
     &         & 24404603 & 1996-07-18 \, 06:29:49& L02 &  \multicolumn{4}{l}{(off-source reference measurement for TDT 24402202)}  \\
     &         & 80500101 & 1998-01-28 \, 10:36:25 & L02 &2.566&2.549&22.2& \\
     &         & 80500104 & 1998-01-28 \, 11:02:50 & L02 &2.566&2.549&22.2& \\
     &         & 80500107 & 1998-01-28 \, 11:29:15 & L02 &2.567&2.549&22.2& \\
     &         & 80500110 & 1998-01-28 \, 11:55:40 & L02 &2.567&2.549&22.2& \\
     &         & 80500113 & 1998-01-28 \, 12:22:05 & L02&2.567&2.549&22.2 & \\
     &         & 80500116 & 1998-01-28 \, 12:48:30 & L02 &2.567&2.549&22.2& \\
     &         & 80500119 & 1998-01-28 \, 13:14:55 & L02 &2.568&2.549&22.2& \\
     &         & 80500122 & 1998-01-28 \, 13:41:20 & L02 &2.568&2.549&22.2& \\
     &         & 80500125 & 1998-01-28 \, 14:07:45 & L02&2.568&2.549&22.2 & \\
     &         & 80500128 & 1998-01-28 \, 14:34:10 & L02 &2.568&2.549&22.2& \\
     &         & 80500131 & 1998-01-28 \, 15:00:35 & L02 &2.569&2.549&22.2& \\
     &         & 80500134 & 1998-01-28 \, 15:27:00 & L02 &2.569&2.549&22.2& \\
     &         & 80500137 & 1998-01-28 \, 15:53:25 & L02 &2.569&2.549&22.2& \\

    \noalign{\smallskip}
(10) & Hygiea  & 83201702 & 1998-02-24 \, 22:08:03 & L01 &3.179&3.444&16.6& CII background emission \\
     &         & 83201803 & 1998-02-24 \, 22:44:05 & L02 &3.179&3.444&16.6& \\
     &         & 84801302 & 1998-03-12 \, 16:13:06 & L02 &3.402&3.434&16.7& \\
     &         & 85303402 & 1998-03-17 \, 16:09:46 & L02 &3.472&3.430&16.6& \\
  \end{longtable}
  \end{center}}
  
\longtab{2}{
\begin{longtable}{ll l l l l l l l}
      \caption{
               \label{tbl:nonstd} 
               Overview of L99 asteroid observations in our catalogue. All observations
               are L01-like scans covering the full wavelength range.}\\

    \hline\hline
    \noalign{\smallskip}
      \multicolumn{2}{l}{Object} & TDT & {Mid-observation UTC}& $\Delta$ $[$AU$]$ & r $[$AU$]$ & $\alpha$ $[\degr]$& {Remarks}\\
     \noalign{\smallskip}
    \hline
    \noalign{\smallskip}
    \endfirsthead
   \caption[]{(continued)}\\
    \hline\hline
    \noalign{\smallskip}
       \multicolumn{2}{l}{Object} & TDT & {Mid-observation UTC}& $\Delta$ $[$AU$]$ & r $[$AU$]$ & $\alpha$ $[\degr]$& {Remarks}\\
     \noalign{\smallskip}
    \hline
    \noalign{\smallskip}
   \endhead
   \noalign{\smallskip}
   \hline
   \endfoot
(1) & Ceres & 07500601 & 1996-01-31 \, 07:52:11 & 2.891 & 2.682 & 19.9 & CII background emission\\
    &	    & 07500701 & 1996-01-31 \, 10:10:56&  2.890 & 2.682 & 19.9 & CII background emission\\ 
    &	    & 07501301 & 1996-01-31 \, 16:08:13 &  2.887 & 2.682 & 19.9 & CII background emission,\\
    &	    &	       &	    &	    &&&non-standard bias voltages \\     
    &	    & 07501401 & 1996-01-31 \, 18:32:03 &  2.886 & 2.682 & 20.0 & CII background emission,\\        
    &	    &	       &	    &	   & &&non-standard bias voltages \\
     \noalign{\smallskip}   
(4) & Vesta & 05500201 & 1996-01-11 \, 07:10:35 &  2.313 & 2.250 & 24.8 &non-standard bias voltages\\      
    &	    & 05500301 & 1996-01-11 \, 08:54:29 & 2.312 & 2.250 & 24.9 & non-standard bias voltages \\
    &	    & 05500401 & 1996-01-11 \, 10:39:34  & 2.311 & 2.250 & 24.9 &non-standard bias voltages \\
  \end{longtable}}

\end{document}